\pdfoutput=1

\documentclass[aps,prl,twocolumn,groupedaddress,amsmath,amssymb]{revtex4-1}
\usepackage{graphicx}

\begin{document}


\title{Nonmonotonic Aging and Memory in a Frictional Interface}


\author{Sam Dillavou$^1$, Shmuel M. Rubinstein$^2$}
\affiliation{$^1$Physics, Harvard University, Cambridge, Massachusetts 02138, USA \\
$^2$Applied Physics, Harvard University, Cambridge, Massachusetts 02138, USA}


\date{\today}

\begin{abstract}
We measure the static frictional resistance and the real area of contact between two solid blocks subjected to a normal load. We show that following a two-step change in the normal load the system exhibits nonmonotonic aging and memory effects, two hallmarks of glassy dynamics. These dynamics are strongly influenced by the discrete geometry of the frictional interface, characterized by the attachment and detachment of unique microcontacts. The results are in good agreement with a theoretical model we propose that incorporates this geometry into the framework recently used to describe Kovacs-like relaxation in glasses as well as thermal disordered systems. These results indicate that a frictional interface is a glassy system and strengthen the notion that nonmonotonic relaxation behavior is generic in such systems.
\end{abstract}


\maketitle


Under constant load, the static coefficient of friction of rock \cite{Dieterich:1972ta}, paper \cite{Heslot:1994gd}, metal \cite{Baumberger:2006bq}, and other materials \cite{Berthoud:1999ha, Bocquet:1998wt} grows logarithmically ad infinitum. This aging phenomenon is central to frictional systems ranging from micro-machines \cite{Corwin:2010es} to the earthquake cycle \cite{Scholz:1998aa, Rice:1983vw, Marone:1998wm}, and is described by the Rate and State Friction laws \cite{Dieterich:1979vq, Rice:1983aa, Ruina:1983hh}, where aging is captured by the evolution of a phenomenological state parameter.  Because most solids have microscopically rough surfaces, when two bodies are brought together, their real area of contact is localized to an ensemble of microcontacts, which sets the frictional strength \cite{BowdenTabor,Archard:1957up,Greenwood:1966boa,Persson:2001kz}. Thus, the strengthening of the interface is frequently attributed to a gradual increase of the real area of contact \cite{Dieterich:1994ux, Berthoud:1999ha}; however, recent compelling evidence suggests that such an effect could also result from the strengthening of interfacial bonds \cite{Li:2011gf}.

Recently \cite{Lahini:2017dq}, it was shown that some systems which exhibit slow relaxation and logarithmic aging can also, evolve non-monotonically under static conditions, exhibiting a Kovacs-like memory effect \cite{Kovacs:1969aa}. Several glassy and disordered systems ranging from polymer glasses to crumpled paper \cite{Josserand:2000tu, Lahini:2017dq} exhibit such non-monotonic relaxation following a two-step protocol. Previous observations of de-aging \cite{Rubinstein:2006dt} in real area of contact suggest that frictional interfaces may also belong to this universality class. If so, this may indicate that dynamics of friction exhibit a memory effect and are richer than previously considered; these dynamics cannot be fully captured by Rate and State or any theory with a single degree of freedom.

Here we experimentally demonstrate that a frictional interface can indeed store memory. Using real time optical and mechanical measurements, we observe that under a constant load, both the static coefficient of friction and the real area of contact may evolve non-monotonically. Additionally, in contrast to the prevailing paradigm, the two physical quantities do not always evolve in tandem; in fact, one may grow while the other shrinks. We further show that this discrepancy arises from the non-uniform evolution of the contact surface. We propose a model that generalizes the geometrical descriptions of contact mechanics to include memory effects and the glassy nature of frictional interfaces. 

\begin{figure} 
\includegraphics[width = .5\textwidth]{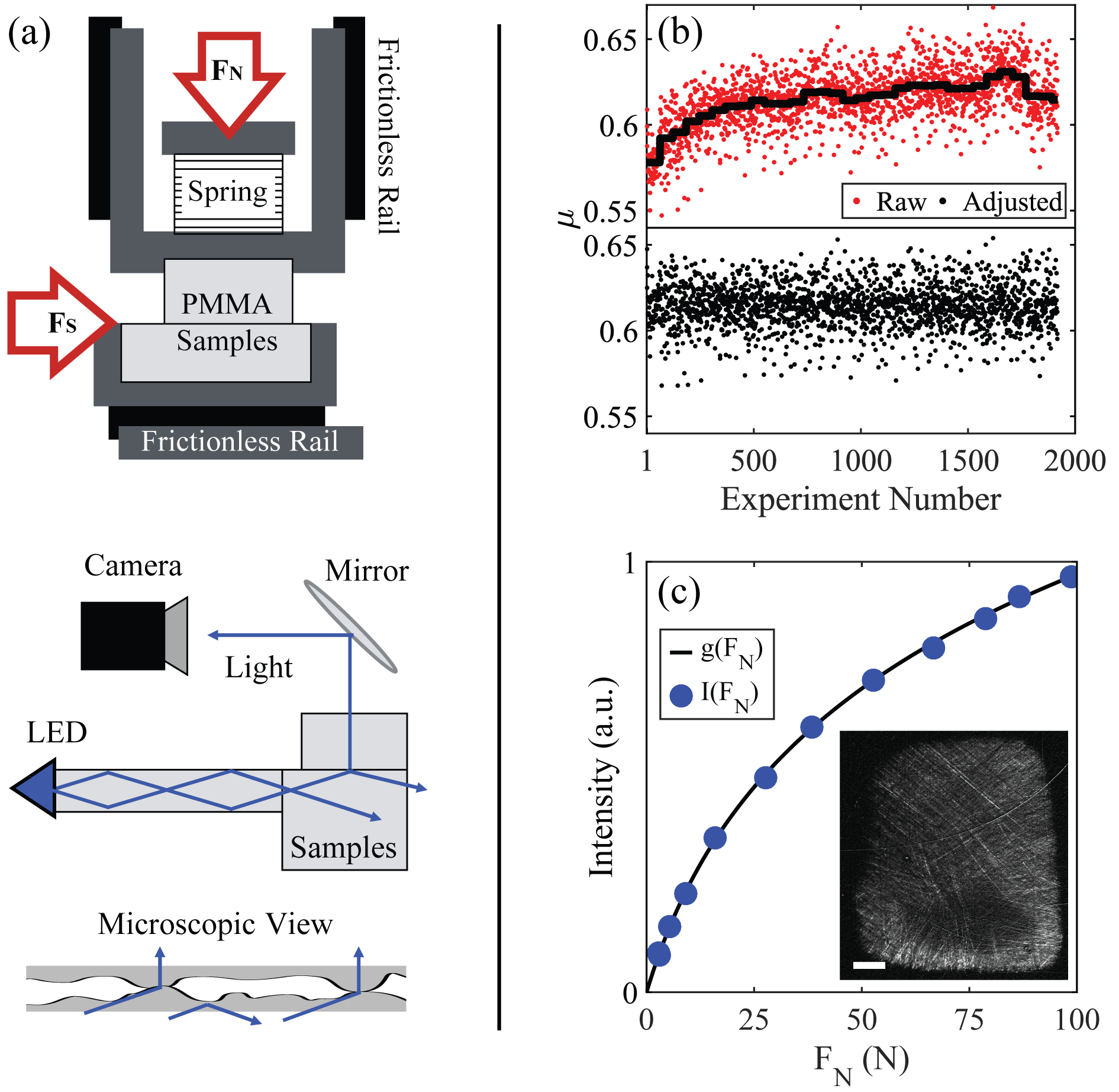}
\caption{Experimental setup
(a) Schematic of the the biaxial compression/translation stage (top) with integrated optical measurement apparatus (bottom)  (b) $\mu_S$ vs experiment number for a typical PMMA-PMMA interface before (top) and after (bottom) trend removal described in text. (c) $I(F_N)$ (blue circles) and $g(F_N)$ (black line) vs $F_N$ for a typical loading cycle. Inset: A typical snapshot of an interface illuminated with TIR at $F_N$ = 100 N after background subtraction. Scale bar is 1 mm.
\label{1}}
\end{figure}

Frictional dynamics are typically described through a force measurement, but understanding the underlying mechanisms requires observation of the 2D interface where shear forces are generated. We thus simultaneously measure the static friction coefficient and real area of contact resolved across an entire interface. Our biaxial compression and translation stage is described schematically in Fig. \ref{1}(a). The interface is formed between two laser-cut PMMA (poly methyl-methacrylate) blocks with 0.5 to 4 cm$^2$ of nominal contact area. Sample surface roughness ranges from the original extruded PMMA ($\sim$50 nm RMS) to surfaces lapped with 220 grit polishing paper ($\sim$50 $\mu$m RMS). A normal load, $F_N$, is applied to the top sample through a spring and load cell, and the bottom sample is held by a frame on a horizontal frictionless translation stage. A shear force, $F_S$, is applied to the bottom frame at the level of the interface by advancing a stiff load cell at a constant rate of 0.1 or 0.33 mm/s. When the load cell makes contact with the frame containing the bottom sample, $F_S$ increases linearly until the interface slips and $F_S$ drops suddenly. We define the coefficient of static friction, $\mu_S$, as the ratio of the peak in shear force to the normal force, $F_P/F_N$. The effect of frictional memory is subtle, and observing it requires many runs, systematically varying multiple parameters. However, sliding wears the samples, which causes $\mu_S$ to vary systematically over the course of many measurements, as shown by the raw data (red) in Fig. \ref{1}(b). In order to differentiate the system's response to a change in the experimental parameters from its slow, background evolution due to wear, we implement a randomization protocol. We test all points of interest in our parameter space once in a random order, then again in a different random order, and so on, such that every point is visited a minimum of 25 times. Additionally, all cycles are normalized to have the same mean, as shown by the adjusted data (black) in Fig. \ref{1}(b). Finally, to minimize the uncertainty associated with initial placement and loading of the samples \cite{BenDavid:2011di, Yamaguchi:2016ga}, we follow two pre-stress protocols \cite{reproducibilityNote}.

The real area of contact between the blocks, $A_R$, is measured by using total internal reflection (TIR) \cite{Rubinstein:2004ek, Rubinstein:2006ca, BenDavid:2010kr, Rubinstein:2007gl, Rubinstein:2006dt}. Blue light (473 nm) is incident on the top surface of the bottom sample at an angle below the critical angle for TIR, allowing light to escape into the top sample only through points of contact, as depicted in the bottom of Fig. \ref{1}(a). When imaged through the top sample, the brightness of the interface corresponds to real contacts, as shown in the inset in Fig. \ref{1}(c). When $F_N$ is increased, microcontacts grow in size and number \cite{Dieterich:1994ux}, which allows more light to pass through the interface; thus, the spatially integrated light intensity within the interface, $I(F_N)$, is a smooth and monotonic function of the normal load, as shown in Fig. \ref{1}(d), consistent with previous observations \cite{Rubinstein:2006ca}. The real area of contact evolves in time; however, for sufficiently rapid axial loading, the classical Bowden and Tabor \cite{BowdenTabor} picture holds, and $F_N$ = $A_R\sigma_Y$, where $\sigma_Y$ is the yield stress of the material. Thus, to convert the light intensity to the real area of contact, a conversion function $I = g(F_N) = g(A_R \sigma_Y)$  is fit individually for each experiment. To avoid ambiguity, we only use $I(F_N)$ during the initial rapid loading, faster than 30 N/s. This calibration is used throughout the experiment to convert intensity to real area of contact: $g^{-1}(I(F_N, t))/\sigma_Y = A_R(F_N, t)$.

\begin{figure} 
\includegraphics[width = .5\textwidth]{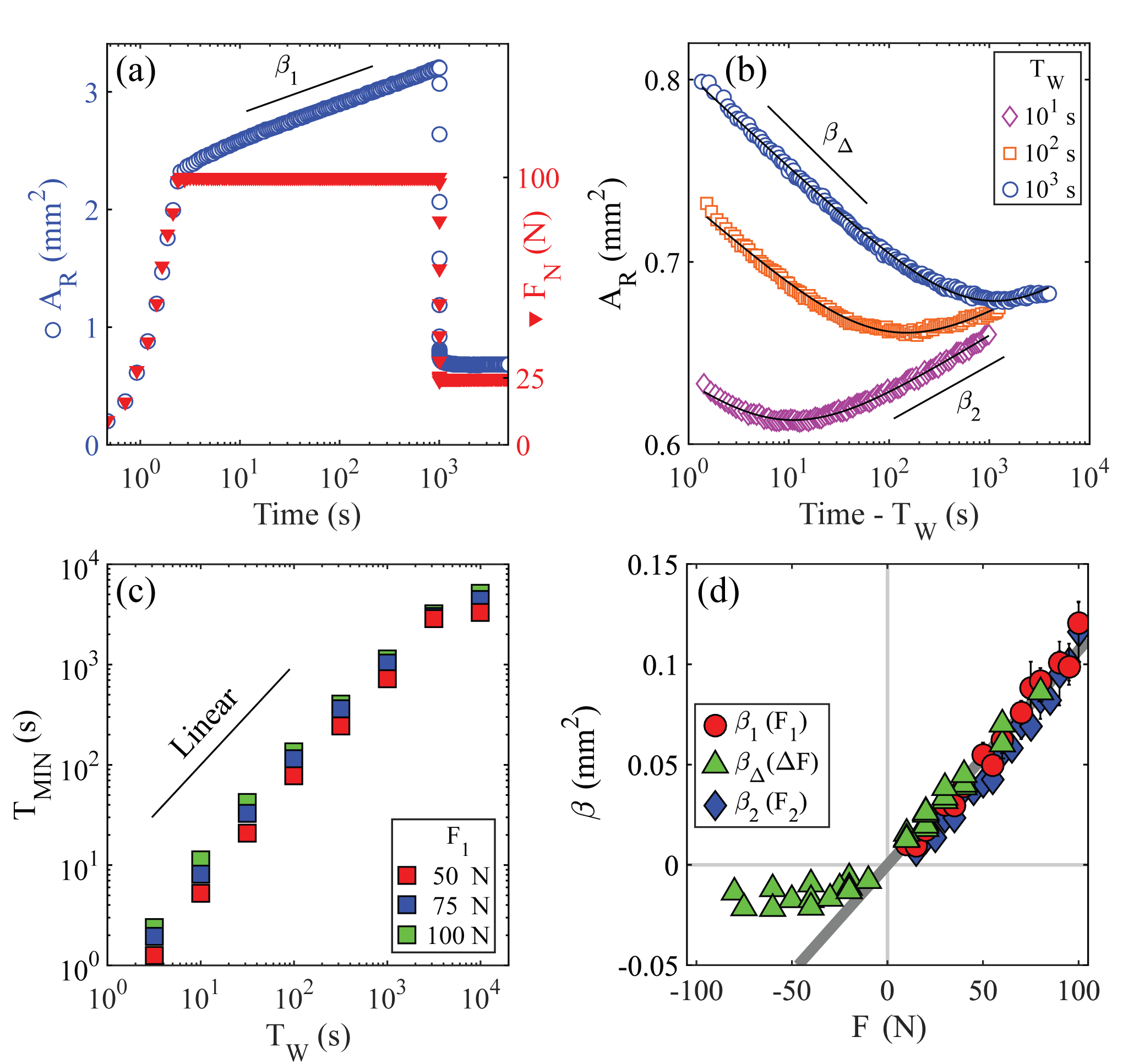}
\caption{Glassy dynamics in the real area of contact
(a) $A_R$ and $F_N$ vs time for a typical two-step protocol with $F_1$ = 100 N, $T_W$ = 1000 s, $F_2$ = 25 N. (b) Evolution of $A_R$ as a function of time shifted by $T_W$ following a step-down in force with $F_1$ = 100 N and $F_2$ = 25 N. Fits (black) use Eq. (\ref{nonmon}). (c) $T_{MIN}$ as a function of $T_W$. $F_2$ = 25 N. (d) $\beta$ vs $F$. $\beta_\Delta$ is plotted vs $\Delta F$. Gray line is a linear fit to $\beta_1$ and $\beta_2$. Note the deviation of de-aging rate  ($\beta_\Delta$, $F<0$) from the linear trend. In (c) and (d), error bars not visible are smaller than data points.
\label{2}}
\end{figure}

We test for memory using a two-step protocol, previously used for other mechanical systems \cite{Lahini:2017dq}, as shown for a typical example in Fig. \ref{2}(a). The blocks are rapidly loaded from above to $F_N$ = $F_1$ and are held constant at that load for a time $T_W$. During this first step, the real area of contact grows logarithmically as \begin{equation} \Delta A_R(t) = \beta_1 \log(t) \end{equation} consistent with previous observations \cite{Dieterich:1994ux, Rubinstein:2006dt}. At t = $T_W$, the normal load is rapidly reduced to $F_N$ = $F_2$ and kept constant for the remainder of the experiment. As a result of the reduction in normal load, many microcontacts instantly detach, showing a simultaneous drop in $A_R$, as shown at $t = 1000$s in Fig. \ref{2}(a). We refer to this instantaneous drop as the elastic response, distinct from the subsequent slow aging. For $t > T_W$  the evolution of $A_R$ is non-monotonic, as shown in Fig. \ref{2}(b). Initially, the real area of contact shrinks in time. This de-aging effect \cite{Rubinstein:2006dt}, or weakening, may persist for seconds, minutes, or even hours until $A_R$ reaches a minimum at some later time, $T_{MIN}$. After this time, $A_R$ increases monotonically, eventually recovering typical logarithmic aging. This is a nontrivial response considering that a nonmonotonic evolution occurs while all loading parameters including $F_N$ are held constant. At any two time points in which $A_R$ has the same value before and after $T_{MIN}$, the load ($F_2$), and all other macroscopic conditions are identical; however, the system's evolution at these two points is opposite in sign. Thus, the non-monotonic behavior clearly indicates that the state of the system cannot be described by a single variable, and additional degrees of freedom storing a memory of the system's history must exist.

The non-monotonic evolution of $A_R$ follows the sum of two logarithms \begin{equation} \Delta A_R (t) = \beta_\Delta \log(t) + (\beta_2 - \beta_\Delta) \log(t+T_W) \label{nonmon}\end{equation} with $\beta_\Delta < 0$, and $T_{MIN} = -\beta_\Delta T_W/ \beta_2$, as shown in Fig. \ref{2}(b). This functional form is consistent with the Amir-Oreg-Imry (AOI) model \cite{Amir:2008et, Amir:2009kc}, recently proposed as a universal model for aging in disordered systems \cite{Lahini:2017dq}. In this framework, the relaxation dynamics of glassy systems are facilitated by a spectrum of uncoupled, exponentially relaxing modes, whose density is inversely proportional to their relaxation time scale. All modes relax to an equilibrium that is set by a control parameter, which for the frictional case is $F_N$. For a two-step protocol, the equilibrium point may change before all modes have fully relaxed; thus, in such a case, non-monotonic evolution may result from fast and slow modes moving in opposite directions. AOI predicts that the evolution of $A_R$ will depend linearly on the parameters in the loading protocol, namely that  \begin{equation} \frac{T_{MIN}}{T_W} = const \quad \quad  \frac{\beta_1}{F_1} = \frac{\beta_\Delta}{\Delta F} = \frac{\beta_2}{F_2} = const \label{scaling}\end{equation}with $\Delta F \equiv F_2-F_1$. These predictions, including the form of Eq. (\ref{nonmon}), also apply to a two-step protocol in which $F_1 < F_2$, albeit with $\beta_\Delta > 0$. Experimentally, we indeed find a linear dependence between $T_{MIN}$ and $T_W$ over several orders of magnitude, as shown in Fig. \ref{2}(c). Such a proportionality is a hallmark of real aging and memory \cite{Lahini:2017dq}. We also find that a step-up protocol ($F_1<F_2$) yields the same double logarithm evolution as the step-down protocol ($F_2<F_1$), and the logarithmic slopes in both protocols are consistent with Eq. (\ref{scaling}), with the notable exception of de-aging, as shown in Fig. \ref{2}(d). We return to this later to discuss a possible resolution to this discrepancy, via a modification of AOI that accounts for the instantaneous detachment of microcontacts that results from a drop in $F_N$.

The non-monotonic effects described above indicate that interfacial memory influences the evolution of the real area of contact. The correspondence between the real area of contact and the static coefficient of friction has been well established \cite{BowdenTabor, Dieterich:1994ux,Rubinstein:2004ek, Rubinstein:2006ca, BenDavid:2010kr, Rubinstein:2007gl, Rubinstein:2006dt,Svetlizky:2014gs,Rubinstein:2009gt}. However, for aging, the vast majority of these tests relied on a single-step protocol, which shows only continuous logarithmic strengthening, captured well by the Rate and State theory. The non-monotonic relaxation we observe in $A_R$ cannot be captured by any single degree of freedom model, including Rate and State; thus, it is important to test if $\mu_S$ also exhibits memory.

\begin{figure} 
\includegraphics[width = .5\textwidth]{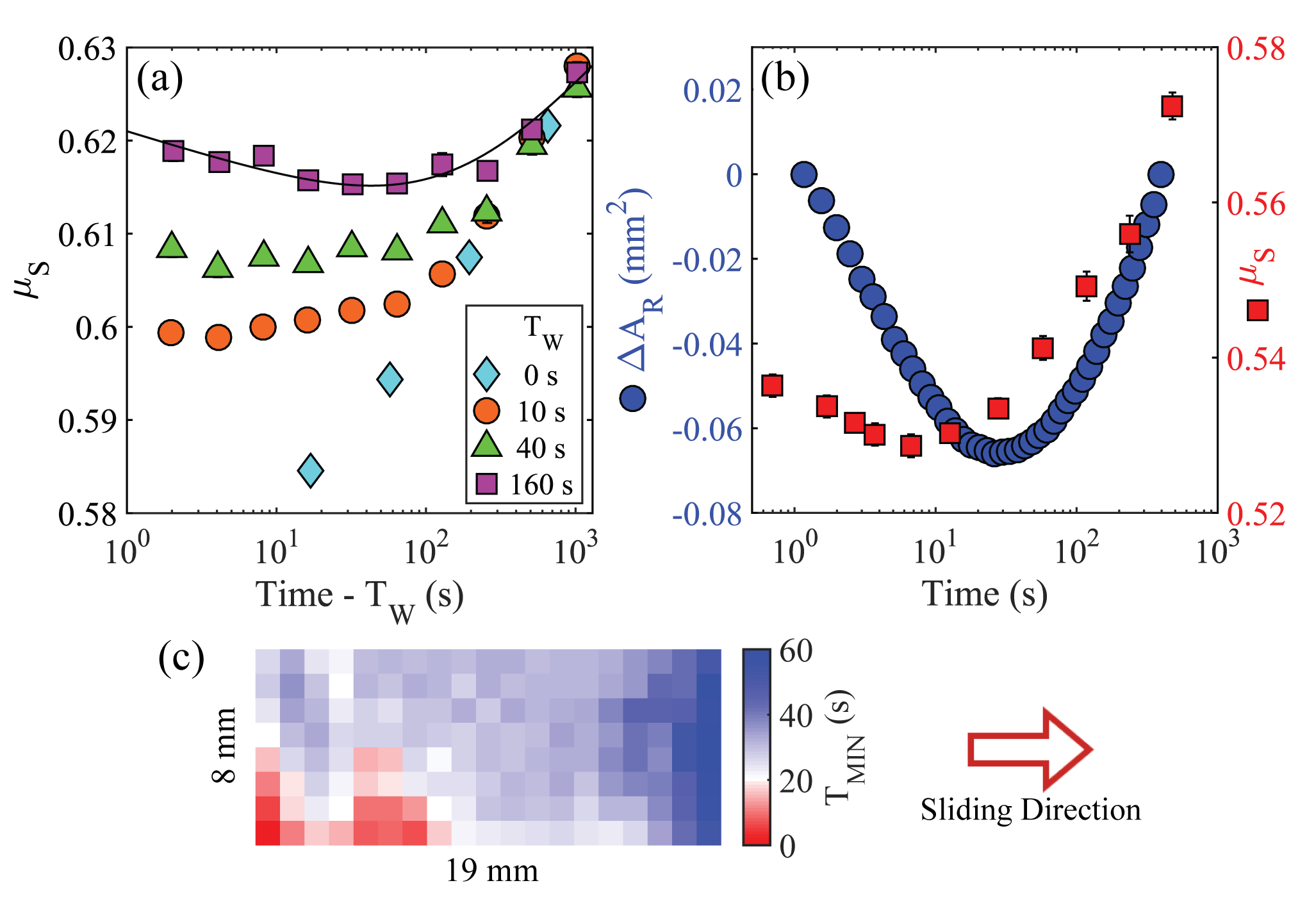}
\caption{Memory in static friction
(a) $\mu_S$ vs time for $F_1$ = 90 N, $F_2$ = 25 N. Line is a guide for the eye to highlight nonmonotonicity.  (b) Evolution of $A_R$ and $\mu_S$ as a function of time shifted by $T_W$ following a step-down in force with $F_1$ = 90 N, $T_W$ = 60 s, $F_2$ = 40 N. (c) Local $T_{MIN}$ of $A_R(x,y)$ for the experiment shown in (b). Notice that $T_{MIN}$ of $A_R(x,y)$ corresponds to $T_{MIN}$ of $\mu_S$ only over a portion near the trailing edge. In (a) and (b), error bars not visible are smaller than data points.
\label{3}}
\end{figure}

Every measurement of $\mu_S$ necessitates slip which resets the interface and the experiment. Therefore, while the full evolution of $A_R$ can be continuously measured in a single experiment, the nonmonotonic behavior of the frictional response cannot be verified in a single run, and measuring $\mu_S(t)$ requires numerous repetitions of any single protocol. Comparing $\mu_S(t)$, as shown in Fig. \ref{3}(a), to $A_R(t)$ reveals that the two physical quantities exhibit a qualitatively similar memory effect, including non-monotonicity and the increase of $T_{MIN}$ with $T_W$. This indicates that a frictional interface is glassy, and can exhibit a real, Kovacs-like memory effect \cite{Lahini:2017dq}. Following a two-step protocol, $\mu_S(t)$ and $A_R(t)$ both evolve non-monotonically, yet they do not evolve synchronously. Concurrent measurements of the two quantities show that for an extended period, $\mu_S(t)$ increases whereas $A_R(t)$ continues to decrease, as evidenced by the time period of 8 to 32 seconds in Fig. \ref{3}(b). This result points to a simultaneous departure from the Bowden and Tabor framework, as well as from Rate and State Friction.

The discrepancy between $\mu_S(t)$ and $A_R(t)$ emerges from the complex nature of the spatially extended, 2D interface. Even for carefully prepared surfaces, loading is never perfectly homogenous \cite{Rubinstein:2007gl}. As a result, the interface displays a plethora of local responses to a two-step protocol, and $T_{MIN}$ can vary significantly across the interface, as shown in Fig. \ref{3}(c). In only a few regions does $A_R(t,x,y)$ shrink and grow in concert with $\mu_S(t)$; less than 15\% of the interface has a $T_{MIN}$ value closer to $\mu_S(t)$ than to $A_R(t)$. This indicates that $A_R(t) = \iint A_R(t,x,y)dxdy$ does not fully represent the state of the interface.

We have shown that both $A_R(t)$ and $\mu_S(t)$ display glassy memory and non-monotonic evolution in time. This behavior cannot be reconciled with a single degree of freedom model like Rate and State Friction, but instead requires a larger spectrum of relaxation modes, such as AOI. Furthermore, we observe $A_R(t)$ and $\mu_S(t)$ evolving asynchronously, and find a dramatic variation in the evolution of $A_R(t,x,y)$ across the interface due to heterogeneity. This variation may explain the inconsistency. It is well accepted that extended bodies with many contact points do not begin sliding uniformly; rather sliding is nucleated within a small region before rapidly propagating outward in a coherent fracture front \cite{Svetlizky:2014gs, Rubinstein:2004ek, Rubinstein:2006ca, Rubinstein:2007gl}. In our system, nucleation occurs near the trailing edge of the top sample \cite{Rubinstein:2004ek}. Therefore, we expect the evolution of $\mu_S$ to be dominated by the evolution of $A_R(x,y)$ in that region. Indeed, near the trailing edge, the value of $T_{MIN}$ for $\mu_S$ and for $A_R(x,y)$ match quite well, as shown on the left of Fig. \ref{3}(c). 

\begin{figure} 
\includegraphics[width = .5\textwidth]{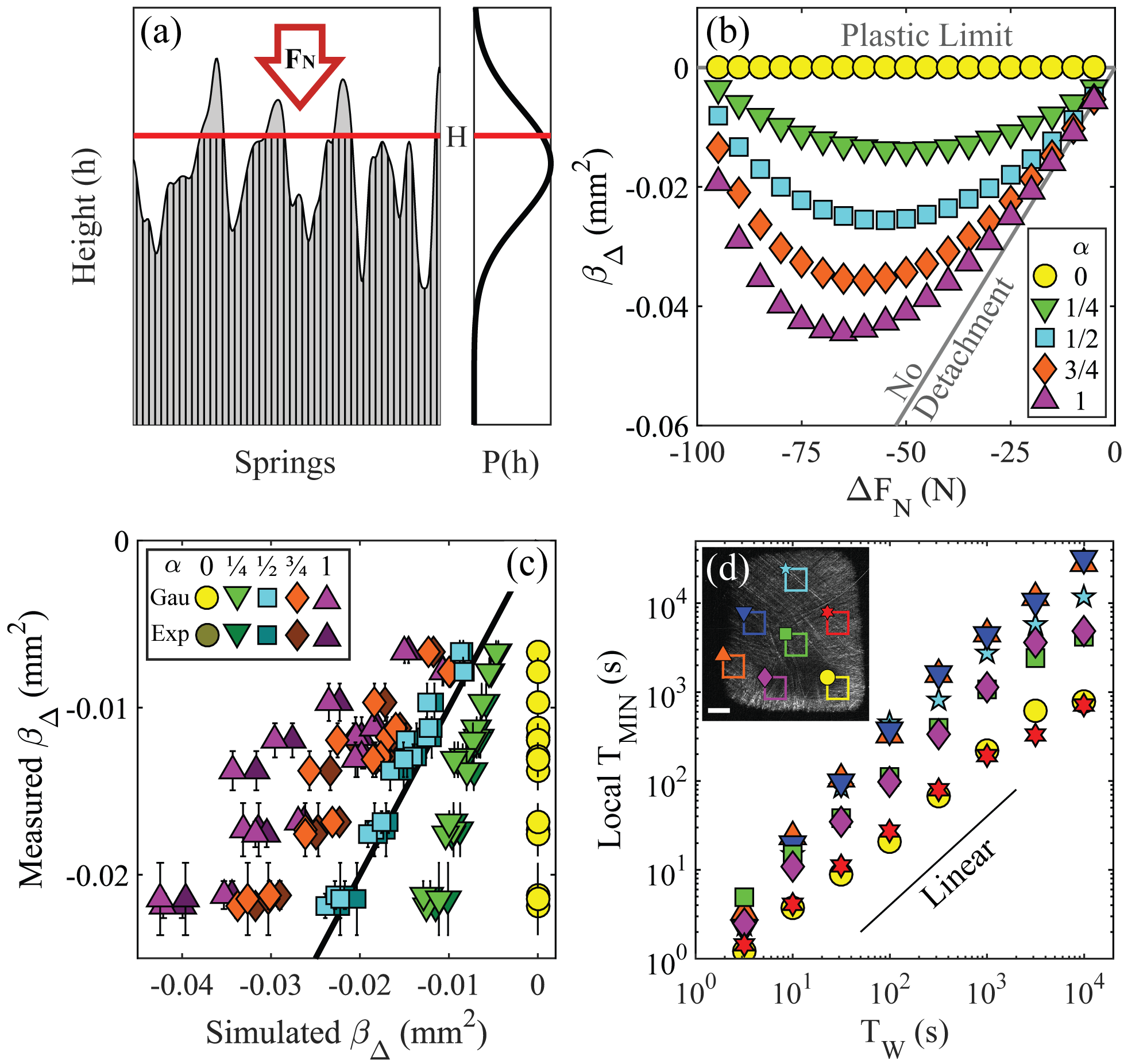}
\caption{A phenomenological model for aging of a frictional interface
(a) Graphical representation of the ensemble of spring-like modes which compose the interface in the model. A rigid line at global height $H(F_N)$ compresses all springs in contact. Probability of a spring height $P(h)$ is shown on the right. (b) Simulated $\beta_\Delta$ vs $\Delta F$ for $F_1$ = 100 N and a Gaussian distribution of heights, $P(h)$. (c) Measured $\beta_\Delta$ vs simulated $\beta_\Delta$ for five values of $\alpha$. A perfect correspondence would lie on the black, $x=y$ line. Darker and brighter shades correspond to exponential and Gaussian distributions of $P(h)$ respectively. (d) Local $T_{MIN}$ of $A_R(x,y)$ vs $T_W$ for $F_1$ = 100 N, $F_2$ = 25 N. The seven locations are indicated in the inset with corresponding colors.  Scale bar is 1 mm. In (c) and (d), error bars not visible are smaller than data points.
\label{4}}
\end{figure} 

Taking into account another heterogeneity of the interface may also suggest a resolution to the anomalously weak de-aging rate in the global $A_R(t)$, following the step-down protocol. As previously noted by Greenwood and Williamson \cite{Greenwood:1966boa}, when $F_N$ is reduced, the separation between the two surfaces increases, and many microcontacts instantly detach \cite{Dieterich:1994ux, Archard:1957up}. Any memory stored in a detached microcontact cannot influence the future evolution of $A_R$. This suggests a generalization of AOI in which each individual mode, $i$, can engage and disengage at a cutoff height, $h_i$, uncorrelated with its time constant, $\lambda_i$. As a result, instead of a single global equilibrium $F_N$, each mode's equilibrium is a function $f(h_i-H)$, where $H(F_N)$ is a global parameter. An appealing interpretation of this model considers an ensemble of springs with spring function $f(h_i-H) = k(h_i-H)^\alpha$ for $h_i\geq H$, and $0$ for $h_i<H$, compressed from above by a rigid, flat plane under force $F_N$,  as shown in Fig. \ref{4}(a). We follow Greenwood and Williamson \cite{Greenwood:1966boa} and assume a normal or an exponential distribution of $h_i$'s. Detachment is introduced by stipulating that modes with $h_i<H$ are disregarded. Including detachment has no effect on asymptotic aging ($\beta_1$ and $\beta_2$), or on positive transient aging ($\beta_\Delta$ for $\Delta F > 0$), but it dramatically reduces the rate of de-aging ($\beta_\Delta$ for $\Delta F < 0$). We find results are insensitive to the probability distribution of spring heights, $P(h)$, provided they are sufficiently broad \cite{Persson:2001kz}. We fit the spring constant, $k$, to match asymptotic aging data, $\beta_1$ and $\beta_2$, leaving a single free parameter in the model, $\alpha$.

The modes can be interpreted as elements of real contact area, in which case $\alpha=0$ generates the fully plastic, Bowden and Tabor picture \cite{BowdenTabor} where all area in contact carries a set pressure (the yield stress) regardless of normal force. Thus, for $\alpha=0$ de-aging is completely eliminated, as shown in Fig. \ref{4}(b). Correspondingly, $\alpha = 1$ describes an ensemble of Hookean springs whose intitial deformation corresponds to fully elastic interfacial models \cite{Greenwood:1966boa,Archard:1957up}. The experimental data matches quite well with $\alpha = 1/2$, falling exactly between the fully plastic, $\alpha =0$ and the fully elastic $\alpha =1$ limits, as shown in Fig. \ref{4}(c). One implication of this model is that rich, non-monotonic aging behavior and the memory effect are not only global properties, but should persist in small subsections of the interface. Indeed, logarithmic aging, de-aging, and the (quasi) linear scaling of $T_{MIN}$ with $T_W$ are also present locally, as shown in Fig. \ref{4}(d). It is natural to wonder onto what small scale the kovacs-like memory effect will persist and whether it could be observed even on a single asperity level.

\bibliography{frictionKovacsbib}

\begin{thebibliography}{33}
\bibitem [{rep()}]{reproducibilityNote}%
  \BibitemOpen
  \href@noop {} {\enquote {\bibinfo {title} {{In the experiments shown in Fig.
  \ref{3}(b), samples are briefly slid at $F_N$ = 30 N prior to performing each
  two-step protocol. In the experiments shown in Fig. \ref{3}(a), prior to each
  experiment $F_N$ is briefly, ($<2$ s), increased to 95 N then held at 3 N for
  30 s. Without these initialization methods, measurements of $\mu_S$ were
  inconsistent.}}}\ }\BibitemShut {NoStop}%
\end{thebibliography}

\end{document}